\documentclass[acmtog]{acmart}

\AtBeginDocument{%
 \providecommand\BibTeX{{%
  \normalfont B\kern-0.5em{\scshape i\kern-0.25em b}\kern-0.8em\TeX}}}

\setcopyright{none}

\acmConference[RecSys'22]{}{September 18--23,
 2022}{Seattle, WA}
\usepackage{multirow}

\begin{document}

\title{Solutions to preference manipulation in recommender systems require knowledge of meta-preferences}

\author{Hal Ashton}
\author{Matija Franklin}
\email{matija.franklin@ucl.ac.uk}
\authornote{Both authors contributed equally to this research.}
\affiliation{%
 \institution{University College London}
 \country{UK}
}

\renewcommand{\shortauthors}{Ashton and Franklin}

\begin{abstract}
Iterative machine learning algorithms used to power recommender systems often change people's preferences by trying to learn them. Further a recommender can better predict what a user will do by making its users more predictable. Some preference changes on the part of the user are self-induced and desired whether the recommender caused them or not. This paper proposes that solutions to preference manipulation in recommender systems must take into account certain meta-preferences (preferences over another preference) in order to respect the autonomy of the user and not be manipulative. 
\end{abstract}




\maketitle

\section{Introduction}

Recommender systems try to predict a user's preferences in order to offer them a personalized selection of options on a given platform. In economics, preference is often defined as a choice between alternative options originating from utility theory. However, this is an operational definition - a definition for a concept which cannot be directly measured. In truth the term preferences is unconscionably vague. Sometimes it covers the user's beliefs about the world, their hedonic state or mood or their psychometric attributes. All this in addition to an ordering over alternatives in the spirit of its use in utility theory. Now is not the space to separate all these meanings out, but we are aware that a finer grain definition and taxonomy of preferences is desirable. 
In line with \citet{franklin2022recognising}, we conceptually define preferences as \emph{any explicit, conscious, and reflective or implicit, unconscious, and automatic mental process that brings about a sense of liking or disliking for something}. In practice we note that in research and practice, the term preferences is also used to describe a person's beliefs or knowledge about the world. The existing discussion of preference manipulation often therefore also concerns the manipulation of beliefs. The provision of a finer set of definitions and a taxonomy is a project for the future. For now the vagueness is tolerable.

In this paper we discuss this a potential solution to preference manipulation in recommender systems; namely, learning meta-preference - people's preferences over their own preferences. After identifying the source of preference manipulation and introducing meta preferences this paper propose a requirement to solutions which claim to solve the recommender preference manipulation problem, itself an example of auto-induced distributional shift \cite{krueger_hidden_2020}. 

\section{The problem of behaviour and preference manipulation in recommender systems}

Machine learning (ML) can be used to understand user preferences in order to improve recommender systems. Recommender systems gather stated or revealed preferences in the attempt to better predict what a user will do in the future. In economics, a choice of one option over another indicates a preference for it. This is an example of a `revealed preference' -- the preference is assumed to be revealed through the agent's actions \cite{samuelson1938note}. `Stated preferences' are the preferences of an agent that are elicited by directly asking the agent \cite{kroes1988stated}. Both approaches have their drawbacks. People are not be fully rational \cite{dhami2022bounded}, they may not accurately report their preferences \cite{sunstein2018unleashed}, and they may be influenced by how questions are worded \cite{vspecian2019precarious}. Behaviour is restricted by the choices open to a user at any time, potentially confounding analysis. 

Over and above this, human preferences change over time and can be manipulated \cite{franklin2022recognising}. If an iterative machine learning algorithm is used, it becomes difficult to know whether the recommender system has learned about its users, whether the users have changed, or whether the system has taught users to behave in ways that maximizes the objective function \cite{ashton2022problem}. This becomes more difficult because of the bidirectional causal relationship between behaviour and preferences \citet{ariely_how_2008}. A recommender system trained to maximise user 'engagement' has an incentive to change users' preferences \cite{Everitt2021AgentPerspective}. A recommender can better do its job in predicting what its users will do if it makes its users more predictable \cite{russell_human_2019}. This has been shown in simulation by, amongst others, \citet{evans_user_2021} and \citet{jiang_degenerate_2019}.

\section{Initial solutions to the manipulation problem}
Various potential solutions have been put forward to solve the preference manipulation problem. \citet{Everitt2021AgentPerspective} show that a recommender serving content to a user using a preference set based on the counterfactual world where the user had not interacted with the system removes the preference manipulative incentive. \citet{carroll_estimating_2022} penalise a recommender system for shifting the preferences of a user 'unnaturally'. The reasoning is promising - penalising recommender systems for the wrong kind of shifts - but the execution is lacking. They pragmatically define natural preference shifts as those that a user would undergo if presented with a random slate of recommendations. \cite{farquhar_path-specific_2022} examine the causal path that may cause preference change and prevent the algorithm from using elements on that path as a means to an end - ie shifting preferences to suit its preferences.

We argue that these solutions are unsatisfactory because they fail to take into account the user's meta-preferences. Some users like the preference shift that content engenders. Here the recommender is a cause of their preference shift but it is a welcome one. The only way to disentangle welcome shifts from unwelcome or even manipulative ones is to learn about the user's meta preferences. Causal analysis alone won't cut it.

\section{Meta preferences}
A meta preference is a preference of another preference and is itself a preference \cite{franklin2022recognising}\footnote{Arbitrary high levels of meta preference are allowable therefore.}. To give a concrete example, someone might enjoy a certain film, but simultaneously wish that they didn't. This we might characterise as a guilty pleasure (see Table 1.). Similarly they might dislike exercise and dislike that about themselves (a guilty chore or hatred perhaps). Further, they may also like that they like certain things (righteous-pleasure) and like that the dislike other things (righteous-hatred). 

Although meta-preferences can be changed, they are more stable than preferences \cite{pettigrew2019choosing}, which can change from moment to moment. Some have called these more fundamental changes to a persons meta-preferences \textit{Transformative Experiences} \cite{paul2014transformative, pettigrew2020transformative}. The presence of a fast-food restaurant may change one's preference for what they want to eat right now, but will not change their meta preference towards being healthy. An chronic illness, on the other hand, can be a Transformative Experience that influences a person's meta-preferences \cite{carel2017illness, hole2020illness}. Do meta preferences thus allow better recommendations? Do people who like something as a guilty pleasure like different things from someone who just flat out likes the thing?

 \begin{table}[]
 \centering
 \resizebox{0.45\textwidth}{!}{%
 \begin{tabular}{|l|ll|}
 \hline
 \multirow{2}{*}{\textbf{Meta  Preference}} & \multicolumn{2}{l|}{\textbf{Preferences}} \\ \cline{2-3} 
  & \multicolumn{1}{l|}{\textbf{Likes}} & \textbf{Dislikes} \\ \hline
 \textbf{Likes} & \multicolumn{1}{l|}{'Righteous-pleasure'} & 'Righteous-hatred' \\ \hline
 \textbf{Dislikes} & \multicolumn{1}{l|}{'Guilty-pleasure'} & 'Guilty-hatred' \\ \hline
 \end{tabular}%
 }
 \caption{Meta-preferences are preferences over preferences}
 \label{tab:my-table}
 \end{table}

A different view of meta-preferences is to consider their intertemporal nature. That is to say the preference over the their preference in the future - \textit{Conative Preferences}. Consider one of the author's dislike for mushrooms and their desire that they liked them more, because they feel that they are missing out on mycological cuisine. An example of this comes from \textit{Free Traits} - people's ability to adopt new personally traits for a given situation they are in \citet{little2008personal}. People often adopt free traits for the purpose of fulfilling a personal project to the best of their ability \cite{little201714}. \citet{little2017prompt} who developed \textit{Free Trait Theory} often gives the autobiographic example of an introverted professor adopting \textit{extraversion} to achieve his personal project of being a good teacher.

The difference between meta-preferences and conative-preferences seems slight but the possession of a guilty pleasure does not imply that the owner wants to to not have it in the future. The manipulation of preferences concerns \textit{caused} changes of user preferences, so its study will involve conative-preferences to some degree. If a recommender system were to cause some change in user preferences against that user's conative-preferences, then that would be a bad outcome. Additionally if a recommender system were to deny someone from satisfying their conative-preferences that would also be bad, potentially trapping users in a unwanted preference state.

\section{Preference-change preferences}

As well as having a preference as to how their preferences evolve in the future, people will also have a preference about what causal mechanism alters their preferences. Previous research has termed these Preference-Change Preferences \cite{franklin2022recognising}. Self-development could be characterised as a voluntary or self-initiated change in preferences through the collection of knowledge or experience. Generally people have a positive preference towards the process. Manipulation might be described as the deliberate change of preferences by an external party. There are those who require manipulation to be hidden from the manipulated \cite{susser_online_2018} but we side with the account of \cite{benn_whats_2022} who do not require this. It is possible to be manipulated even if we are aware of it and how it works. Unlike self-development, manipulation is not done with the permission of the target and is autonomy reducing. Ethically, algorithmic manipulation is dubious \cite{christiano_algorithms_2022}, and occasionally illegal as the EU AI Act details \cite{franklin_missing_2022}. The ideal recommender system would respect the user's preference-change preferences.

\section{Preference-change consent}
If a recommender system were to understand how any recommendation were to change the preferences of a user, it would be able to ask that user, before serving it, whether the user was comfortable with that. By doing so, the recommender would be able to respect the meta-preferences of the user. 

Looking around in the real world, one can see examples of this in the film classification system which warns viewers about the content of films, \textit{before} they watch them. Whilst they do not say what effect a film will have on a viewer, by warning the viewer about elements of the film, they will often make it possible for a viewer to avoids films that elicit certain reactions.

Would this end manipulation by recommender systems? Unfortunately not. Whilst preference manipulation will be made harder if a recommender system obeys a users meta-preferences, it might not make it impossible. The recommender could aim to change the user's meta-preferences and thereby legitimately change their preferences. The ease of achieving this is unknown and a suitable question for further investigation. 

Recommender engines could additionally tell a user what they currently think a user's preferences are. Such an interaction would be possible with the use of example based Explainable Artificial Intelligence (XAI) methods \cite{van2021evaluating}. Example-based explanations provide users historical situations (e.g., films that they have watched) to the current situation (e.g., film that is being recommended). This would serve as a meta preference elicitation method. A user could choose whether or not the learnt preference does in fact match their preferences more broadly - their meta-preferences. For example a recommender system may learn from a users' behaviour that they have a preference for 1970s musicals. A participant can disclose whether or not this does match their preferences. If not the recommender engine can treat it as an anomaly rather than as a new drastic change in the users preference.

\section{Conclusion}

This paper outlines the the preference manipulation problem in recommender systems and briefly introduces meta-preferences. It is our position that by learning users' meta-preferences, a recommender system can better align with what users want, while being autonomy respecting. Preventing a recommender from causing a user's preferences is not desirable because in many situations a user does consent to them being changed by the recommender. Human-in-the-loop architectures have been used to learn from users' preferences \cite{christiano2017deep} and with work can be extended to aid in what we propose.

\begin{acks}
Hal Ashton was supported by a EPSRC doctoral grant.
\end{acks}

\bibliographystyle{ACM-Reference-Format}
\bibliography{sample-base}


\end{document}